\documentclass[11pt, nobibnotes, prl,amsmath]{revtex4}
\usepackage{amsmath, amsthm, amssymb}
\usepackage{psfig}
\usepackage{amsmath, amsthm, amssymb}

\newcommand{\beqn}{\begin{equation}}
\newcommand{\eeqn}{\end{equation}}

\begin{document}

\title{Proof that the Hydrogen-antihydrogen Molecule is Unstable.}

\author{D.K. Gridnev}
\email[Electronic address: ]{dima_gridnev@yahoo.com}
\author{C. Greiner}
\affiliation{Institut f{\"u}r Theoretische Physik,
Robert-Mayer-Str. 8-10, D--60325 Frankfurt am Main, Germany}
\begin{abstract}
In the framework of nonrelativistic quantum mechanics we derive a
necessary condition for four Coulomb charges $(m_{1}^+ , m_{2}^- ,
m_{3}^+ , m_{4}^- )$, where all masses are assumed finite, to form
the stable system. The obtained stability condition is physical
and is expressed through the required minimal ratio of Jacobi
masses. In particular this provides the rigorous proof that the
hydrogen-antihydrogen molecule is unstable. This is the first
result of this sort for four particles.

\end{abstract}


\maketitle

\section{Introduction}\label{sec-intro}

Recent success in the production of trapped antihydrogen atoms
\cite{experiment} has renewed interest in the interaction of
matter with antimatter and especially in the hydrogen-antihydrogen
system (H-$\overline{\text{H}}$). It has long been conjectured
that with pure Coulomb forces no bound state of
hydrogen-antihydrogen exists. The numerical evidence supports this
conjecture \cite{bressani}, yet there is a lack of rigorous proof
as remarked by some authors \cite{armour}, \cite{richard}. Our aim
in this Letter is: (i) to supply such proof under assumption that
only Coulomb forces act between the constituents, (ii) to provide
insight into the screening effect within the system of four
charged particles.

The instability of H-$\overline{\text{H}}$ is explained by the
screening effect just like the instability of the muonic hydrogen
ion (H$\mu^- $). In the system (H$\mu^- $) the heavy muon gets so
tight to the nucleus that screens the positive charge and the
electron ``sees'' a tightly bound neutral combination $p\mu^-$ and
departs from it making the whole system unbound. In \cite{one} we
have proved that the screening effect in the system of three
charged particles is expressed through some critical ratio of
Jacobi masses. From the physical point of view there are two
orbits in this system to consider, namely one orbit within the
pair of particles (the pair that sets up the dissociation
threshold) and the orbit of the third particle in the field of
this pair with respect to the pair's center of mass. Inverse
Jacobi masses are proportional to the Bohr radii of these orbits.
If, say, the orbit of one negative particle is outdistanced then
the attraction from the positive charge is screened by  the other
negative particle and the system becomes unbound.

The system of four unit charges $(m^{+}_1 , m^{-}_2 , m^{+}_3 ,
m^{-}_4 )$ can be unstable only against dissociation into two
neutral pairs. Indeed, if the lowest dissociation threshold would
be dissociation into one particle and the bound cluster of three
particles, then these two objects would have opposite charges and
the long-tailed Coulomb attraction between them would guarantee
the existence of a bound state below the dissociation threshold.
(Just the same argument explains why atoms are stable). This
suggests that we have to consider three orbits, two inner orbits
of the neutral pairs and the third orbit of the relative motion of
these pairs. The Jacobi masses for the neutral pairs are $\mu_x =
m_{1}^+ m^{-}_2 / (m^{+}_1 + m^{-}_2 )$ and $\mu_y = m^{+}_3
m^{-}_4 / (m^{+}_3 + m^{-}_4 )$. The Jacobi mass corresponding to
the relative motion of these two pairs is $\mu_R = (m^{+}_1 +
m^{-}_2 )(m^{+}_3 + m^{-}_4 )/(m^{+}_1 + m^{-}_2 + m^{+}_3 +
m^{-}_4 )$. Pay attention that we order the particles so that
among two possible rearrangements into neutral pairs the lowest
energy threshold corresponds to the dissociation into $(m^{+}_1
m^{-}_2 ) + (m^{+}_3 m^{-}_4 )$ and the pairs are ordered so that
$\mu_x \geq \mu_y$.

The Jacobi masses are in fact not independent, it is easy to check
that $\mu_R \geq 4 \mu_y $ always holds if the particles are
ordered as above. Let us consider the screening effect within the
system of four particles keeping in mind that the Bohr radii of
the orbits are inverse proportional to the Jacobi masses. The
first possibility is $\mu_y \ll \mu_R $ and $\mu_R \approx \mu_x
$. This would mean that three particles form a heavy cluster and
one particle is outdistanced from it. Here everything depends on
whether this heavy three-body cluster has a bound state. If it
does then the whole system is stable because the cluster and the
particle have opposite charges. Hence the whole situation reduces
to the question whether there is any screening in the three-body
system. For $\mu_R \approx \mu_x $ there is no apparent screening
as follows from \cite{one}, \cite{martin1}. For example,
Bressanini {\em et. al.} \cite{bressani} have collected the
convincing evidence that the system $(m^{+}_1 , 1^- , m_{3}^+ ,
1^- )$ is stable for any $m^{+}_1 $ and $ m_{3}^+ $. The
three-body system $(m^{+}_1 , 1^- , 1^- )$ is always stable and if
$ m_{3}^+ \ll 1$ then we run into the situation where $\mu_y \ll
\mu_R$ and $\mu_R \approx \mu_x $ and still the whole system is
stable. This means that $\mu_y \ll \mu_R $ is not sufficient for
the screening effect to take over.

Let us consider the possibility when $\mu_R \ll \mu_x$, which
would mean that the pair $(m^{+}_1 , m^{-}_2 )$ has a very short
inner orbit and other particles are outdistanced from it. In this
case it is right to expect screening because other charged
particles would ``see'' the tightly bound pair $(m^{+}_1 , m^{-}_2
)$ as neutral and the system would fall apart. Our aim in this
Letter is to present the rigorous and analytic proof of this
screening effect, namely
\begin{equation}\label{result}
\mu_R \leq 0.067 \mu_x \: \Longrightarrow \: \text{Instability},
\end{equation}
where under Instability we mean the absence of a bound state below
the dissociation thresholds. Eq.~(\ref{result}) manifests the
screening effect for four particles. From Eq.~(\ref{result}) it
easily follows that the hydrogen-antihydrogen molecule has no
bound state and must decay into protonium and positronium. Muonic
molecules $p \mu^- e^+ e^- $ and $\mu^+ \overline{p} e^+ e^- $ are
unstable as well.

The proof of Eq.~(\ref{result}) is along the same line as in
\cite{one} (the basic idea of the proof is due to Thirring
\cite{thirring}). Before we proceed with the proof let us
introduce the notations. Let $q_i, {\bf r}_i \in \mathbb{R}^3 $
denote charges and position vectors of the particles $i =
1,2,3,4$. We shall work in the system of units where $\hbar = 1$.
We put $q_{1,3} = +1$, and $q_{2,4} = -1$, and the interactions
between the particles are $V_{ik} = q_i q_k / |{\bf r}_i - {\bf
r}_k |$ (remember how the particles are ordered). The stability
problem with Coulomb interactions is invariant with respect to
scaling all masses \cite{martin1}, so we can put $\mu_x =2$. By
the end we shall rescale the masses back.

Consider the system of four charged particles which is stable for
$\mu_R < 3/8$ (this is weaker than in Eq.~(\ref{result})). We
separate the center of mass motion in the Jacobi frame
\cite{messiah} putting ${\bf x} = {\bf r}_2 - {\bf r}_1$, ${\bf y
} = {\bf r}_4 - {\bf r}_3$, ${\bf R} = -a{\bf x} + {\bf r}_3 -
{\bf r}_1 + b{\bf y}$, where $a = m_2 /(m_1 + m_2 )$ and $b = m_4
/(m_3 + m_4 )$ are the mass parameters invariant with respect to
mass scaling. With Jacobi momenta defined as ${\bf p}_{x,y,R} =
-i\nabla_{x,y,R}$ the Hamiltonian of the system takes the form
\begin{equation}\label{ham0}
      H = h_{12} + h_{34}  + \frac{{\bf p}_{R}^2}{2\mu_R} + W,
\end{equation}
where
\begin{equation}\label{W0}
W  = V_{13} + V_{14} + V_{23} + V_{24},
\end{equation}
and $h_{12} = {\bf p}_{x}^2 /4 - 1/x$, and $h_{34} = {\bf p}_{y}^2
/(2\mu_y) - 1/y$ are the Hamiltonians of the pairs (1,2) and (3,4)
(notation $x$ is used instead of $|{\bf x}|$). The ground state
wave function of $h_{12}$ is $\phi_0 = \sqrt{8/ \pi} \exp(-2x)$ so
that $h_{12} \phi_0 = -\phi_0$. By the particle ordering the
energy threshold corresponding to dissociation into two neutral
pairs is $E_{th} = -1 - \mu_y /2$, which is the sum of the binding
energies of the pairs (1,2) and (3,4). Following \cite{one} we
shall cut off the positive part of $W$ by introducing $W_- \equiv
(|W| - W)/2$ and $W_+ \equiv (|W| + W)/2$ which results in the
decomposition $W = W_+ - W_-$, where $W_{\pm} \geq 0$. Instead of
$H$ we shall consider the Hamiltonian
\begin{equation}\label{ham}
    \tilde H = h_{12} + h_{34} + \frac{{\bf p}_{R}^2}{2\mu_R} -
    W_-.
\end{equation}
(The operator $\tilde H$ is self-adjoint on the same domain as
$H$, see \cite{one}). We shall assume that $H $ is stable, {\it
i.e.} $H$ has a bound state $\Psi$ with the energy $E < E_{th}$.
Because $\tilde H \leq H$ we conclude that $\langle \Psi | \tilde
H | \Psi \rangle < E_{th}\| \Psi \|^2 $. Before we use this
inequality let us introduce a projection operator $P_0$, which
acts on any $f({\bf x}, {\bf y}, {\bf R})$ as
\begin{equation}\label{p_0}
    P_0 f \equiv \phi_0 (x) \int d {\bf x}' \phi_0 (x') f({\bf x}' , {\bf y},
{\bf R}),
\end{equation}
and put $\eta = P_0 \Psi$ and $\xi = (1 - P_0 ) \Psi$, where
obviously $\eta \bot \xi$ and $\Psi = \eta + \xi$. We shall assume
that $\| \xi \| \neq 0$ (later we shall get rid of this
assumption), then we are free to choose such normalization of
$\Psi$ that $\| \xi \| = 1$. Now let us rewrite the inequality
$\langle \Psi | \tilde H | \Psi \rangle < E_{th}\| \Psi \|^2 $
decomposing $\Psi$ into $\Psi = \eta + \xi$.
\begin{gather}
  \langle \eta |  h_{34} | \eta \rangle +
  \langle \eta |  \frac{{\bf p}_{R}^2}{2\mu_R} -W_- | \eta \rangle - \langle
  \eta | W_- |  \xi \rangle  - \langle \xi | W_- |
 \eta \rangle \nonumber \\
 + \langle \xi | h_{12} |  \xi \rangle  +  \langle \xi | h_{34}  | \xi \rangle
 - \langle \xi |  \frac{{\bf p}_{R}^2}{2\mu_R} - W_-|  \xi \rangle < -1 -(\mu_y/2) (\|\eta\|^2 + 1) , \label{eq2}
\end{gather}
where we have used that the terms like $\langle \eta | {\bf
p}_{y}^2 | \xi \rangle$ cancel because $P_0$ commutes with the
operators ${\bf p}_{y}^2 , {\bf p}_{R}^2 $ and $1/y$. Indeed, in
this case for example $\langle \eta | {\bf p}_{y}^2 | \xi \rangle
= \langle \eta | P_0 {\bf p}_{y}^2 | \xi \rangle = \langle \eta |
{\bf p}_{y}^2 P_0  | \xi \rangle =0 $.

We are going to rewrite Eq.~(\ref{eq2}) using lower bounds for
some of its terms. From the Hydrogen ground state and by the
variational principle for the terms in Eq.~(\ref{eq2}) the
following inequalities hold $\langle \eta | h_{34} | \eta \rangle
\geq -(\mu_y/2)\|\eta\|^2$ and $\langle \xi | h_{34} | \xi \rangle
\geq -\mu_y/2$. Introducing two non-negative constants $\alpha
=\sqrt{\langle \eta | W_- | \eta \rangle }$ and $\beta =
\sqrt{\langle \xi | W_- | \xi \rangle}$ we get by virtue of the
Schwarz inequality $|\langle \xi | W_- | \eta \rangle | \leq
\alpha \beta$. It remains to figure out the bound for the term $
\langle \xi | h_{12} |  \xi \rangle$. From the bound spectrum of
the Hydrogen atom we have we have \cite{thirring} $h_{12} \geq
-P_0 - 1/4 (1-P_0 )$. (Indeed, $P_0$ projects on the ground state
of $h_{12}$ which has the energy $-1$ and the energy of all other
excited states is greater or equal to $-1/4$ which is the energy
of the second excited state). Hence for the first term in
Eq.~(\ref{eq2}) we get the bound $\langle \xi | h_{12} | \xi
\rangle \geq - 1 /4$. Substituting this into Eq.~(\ref{eq2})
leaves us with the main inequality
\begin{equation}\label{main}
  \langle \eta | \frac{{\bf p}_{R}^2}{2\mu_R} - W_- | \eta\rangle -  2 \alpha
  \beta + \langle \xi |
 \frac{{\bf p}_{R}^2}{2\mu_R} - W_- | \xi\rangle < -\frac 34 .
\end{equation}
We shall focus on the third term on the lhs of Eq.~(\ref{main}).

First, let us prove that the inequality
\begin{equation}\label{f(R)}
\frac{{\bf p}_{R}^2}{2\mu_R} + AV_{14} + AV_{23} \geq  -2A^2 \mu_R
\end{equation}
holds in the operator sense for any $A \geq 0$. Indeed, the
interactions have the form $V_{14} = -1/|\mathbf{R} - \mathbf{z}_1
|$ and $V_{23} = -1/|\mathbf{R} - \mathbf{z}_2 |$, where the
vectors ${\bf z}_1 = -a{\bf x} - (1-b){\bf y}$ and ${\bf z}_2 =
(1-a){\bf x} + b{\bf y}$ play the role of parameters in
Eq.~(\ref{f(R)}). According to the result of Lieb and Simon
\cite{monotone} the energy of the Hamiltonian on the lhs of
Eq.~(\ref{f(R)}) is monotonically increasing with $|\mathbf{z}_1 -
\mathbf{z}_2 |$. Hence the minimum energy is attained when ${\bf
z}_1 = {\bf z}_2 = 0$, which gives us Eq.~(\ref{f(R)}). From
Eq.~(\ref{f(R)}) using the obvious inequality $-W_- \geq V_{14} +
V_{23}$ we find that for any $A \geq 0$ and $\chi({\bf x}, {\bf
y}, {\bf R})$
\begin{equation}\label{l1}
     \langle \chi | \frac{{\bf p}_{R}^2}{2\mu_R} - A W_- | \chi \rangle
     \geq -2 A^2 \mu_R \|\chi\|^2
\end{equation}
holds. With the help of Eq.~(\ref{l1}) we obtain the following
chain of inequalities
\begin{gather}
 \langle \xi | \frac{{\bf p}_{R}^2}{2\mu_R} - W_- | \xi\rangle = \max_{\lambda \geq
 -1} \left[  \langle \xi | \frac{{\bf p}_{R}^2}{2\mu_R} - (\lambda + 1)W_- |
 \xi\rangle + \lambda \beta^2 \right] \geq \label{chain1}\\
\max_{\lambda \geq -1} \left[ - 2(\lambda +
 1)^2 \mu_R + \lambda \beta^2  \right] = \frac{\beta^4 }{8\mu_R} -
 \beta^2,  \label{chain2}
\end{gather}
where in Eq.~(\ref{chain1}) we have added and subtracted the term
$\lambda \beta^2 = \lambda \langle \xi | W_- | \xi\rangle$.
Substituting Eq.~(\ref{chain1})--(\ref{chain2}) into
Eq.~(\ref{main}) leaves us with the inequality
\begin{equation}\label{ratio2}
  \langle \eta | \frac{{\bf p}_{R}^2}{2\mu_R} - W_- | \eta\rangle  + \frac{\beta^4 }{8\mu_R} -
 \beta^2 - 2 \alpha\beta < -\frac 34 .
\end{equation}
The following inequality always holds
\begin{equation}\label{oneline}
\frac{\beta^4 }{8\mu_R} -
 \beta^2 - 2 \alpha\beta  + \frac 34 \geq - \left( \sqrt{\frac{3}{8\mu_R }} - 1 \right)^{-1}
 \alpha^2 .
\end{equation}
To see that Eq.~(\ref{oneline}) is true it suffices to take all
terms to the lhs and minimize over $\alpha$. Substituting
Eq.~(\ref{oneline}) into Eq.~(\ref{ratio2}) and using $\alpha^2 =
\langle \eta | W_- | \eta \rangle$ allows us to formulate the
stability condition
\begin{equation}\label{last}
\langle \eta | \frac{{\bf p}_{R}^2}{2\mu_R} - \left(1 +
(\sqrt{3/(8\mu_R) } - 1)^{-1} \right)W_- | \eta\rangle < 0.
\end{equation}
It remains to consider the case when $\| \xi \| = 0$. It is easy
to see that in this case the substitution $\Psi = \eta$ into the
inequality $\langle \Psi | \tilde H | \Psi \rangle < E_{th} \|
\Psi \|^2 $ leads to the condition even more stringent than
Eq.~(\ref{last}).

It makes sense to introduce the effective potential $V_{eff} ({\bf
y}, {\bf R}) = \int d {\bf x} |\phi_0 |^2 W_- $. The function
$\eta$ has the factorized form $\eta = \phi_0 ({\bf x}) f ({\bf
y}, {\bf R})$. From Eq.~(\ref{last}) we conclude that the system
of four unit charges is unstable if for any fixed ${\bf y}$
\begin{equation}\label{stab1}
\frac{{\bf p}_{R}^2}{2 \mu_R} - \left(1 + ( \sqrt{3/(8 \mu_R) } -
1 )^{-1} \right) V_{eff} \geq 0.
\end{equation}

We shall make one simplification, which helps carrying out all
calculations analytically. We have $W = W_1 + W_2$, where $W_1 =
V_{14} + V_{24}$ and $W_2 = V_{13} + V_{23}$ and obviously $W_-
\leq (W_1 )_- + (W_2 )_- $. Breaking the kinetic energy term in
Eq.~(\ref{stab1}) we deduce that the system is unstable if both of
the following inequalities are satisfied ${\bf p}_{R}^2 - 4 \mu_R
\left( 1 + (\sqrt{3/(8 \mu_R)} - 1 )^{-1} \right) V_{eff}^{(i)}
\geq 0$ for $i = 1,2$, where $V_{eff}^{(i)} = \int d {\bf x}
|\phi_0 |^2 (W_i )_- $. Using the explicit calculation from
\cite{one} we get $V_{eff}^{(1)} \leq (3/16)|{\bf R} + (1-b){\bf
y}|^{-2}$ and $V_{eff}^{(2)} \leq (3/16)| {\bf R} -b {\bf
y}|^{-2}$. It is known \cite{courant} that ${\bf p}_{R}^2 -
\lambda /R^2 \geq 0$ for $\lambda \leq 1/4$. Thus both
inequalities are satisfied if $3 \mu_R \left( 1 + (\sqrt{3/(8
\mu_R)} - 1 )^{-1} \right) \leq 1 $. Solving this inequality and
rescaling the masses tells us that the system is unstable if
$\mu_R \leq (13 - 2 \sqrt{22} ) \mu_x /54$, which proves
Eq.~(\ref{result}). One can improve the constant in
Eq.~(\ref{result}) if one extracts everything from
Eq.~(\ref{stab1}). We preferred to make the simplification by
splitting $W$ into two terms because this makes the whole
derivation analytical. Let us also remark that Instability in
Eq.~(\ref{result}) means that there is no bound state neither
below nor {\em at} the threshold. Indeed, if we would have $H \Psi
= E_{th} \Psi$ then, because one can choose $\Psi > 0$ in the
ground state, we immediately get $\langle \Psi | \tilde H | \Psi
\rangle < E_{th}\| \Psi \|^2 $ which was the starting point of our
analysis.


\begin{thebibliography}{99}

\bibitem{experiment}
M. Amoretti {\em et.al.}, Nature (London) {\bf 419}, 456 (2002);
G.~Gabrielse {\em et.al.}, Phys.~Rev.~Lett.~{\bf 89}, 213401
(2002); G.~Gabrielse  {\em et.al.}, Phys.~Rev.~Lett.~{\bf 89},
233401, (2002)

\bibitem{bressani}
D. Bressanini, M. Mella and G. Morosi, Phys.~Rev.~A~{\bf 55}, 200
(1997)

\bibitem{armour}
E.A.G. Armour and C.~Chamberlain, Few-Body~Systems {\bf 31}, 101
(2002)


\bibitem{richard}
J.M.~Richard, Phys.~Rev.~A~{\bf 49}, 3573 (1994); Few-Body~Systems
{\bf 31}, 107 (2002)

\bibitem{one}
D.K. Gridnev, C.~Greiner and W. Greiner, arXiv math-ph/0502022, to
appear in J.~Math.~Phys. {\bf 46}, issue 4.

\bibitem{martin1}
A. Martin, J.M. Richard and T.T. Wu, Phys.~Rev.~A~{\bf 46}, 3697
(1992)

\bibitem{thirring}
W. Thirring, {\it Lehrbuch der Mathematischen Physik}
(Springer--Verlag, Wien 1994), Vol.~3

\bibitem{messiah}
W. Greiner, {\em Quantum Mechanics: An Introduction}
(Springer--Verlag, Berlin 2000)

\bibitem{monotone}
E. H. Lieb and B. Simon, J.~Phys.~{\bf B11} L537 (1978)


\bibitem{courant}
R. Courant and D. Hilbert, {\em Methods of Mathematical Physics},
Interscience Publishers, New York, (1953), vol.~1, p. 446 ; M.
Reed and B. Simon, {\em Methods of Modern Mathematical Physics},
vol.~2-4, Academic Press (1978)



\end{thebibliography}
\end{document}